# Time-resolved particle image velocimetry measurements with wall shear stress and uncertainty quantification for the FDA benchmark nozzle model.

Jaime S. Raben[1,2], Prasanna Hariharan[2], Ronald Robinson[2], Richard Malinauskas[2] and Pavlos P. Vlachos[3$]

1. School of Biomedical Engineering and Sciences, Virginia Tech
2. Food and Drug Administration, Silver Spring, MD
3. School of Mechanical Engineering, Purdue University

$ communication author : Pavlos P. Vlachos

585 Purdue Mall

Purdue University

School of Mechanical Engineering

pvlachos@purdue.edu

# Abstract

We present validation of benchmark experimental data for computational fluid dynamics (CFD) analyses of medical devices using advanced Particle Image Velocimetry (PIV) processing and post-processing techniques. This work is an extension of a previous FDA-sponsored multi-laboratory study, which used a medical device mimicking geometry referred to as the FDA benchmark nozzle model. Time-resolved PIV analysis was performed in five overlapping regions of the model for Reynolds numbers in the nozzle throat of 500, 2,000, 5,000, and 8,000. Images included a two-fold increase in spatial resolution in comparison to the previous study. Data was processed using ensemble correlation, dynamic range enhancement, and phase correlations to increase signal-to-noise ratios and measurement accuracy, and to resolve flow regions with large velocity ranges and gradients, which is typical of many blood-contacting medical devices. Parameters relevant to device safety, including shear stress at the wall and in bulk flow, were computed using radial basis functions (RBF) to improve accuracy. In-field spatially resolved pressure distributions, Reynolds stresses and energy dissipation rates were computed from PIV measurements. Velocity measurement uncertainty was estimated directly from the PIV correlation plane, and uncertainty analysis for wall shear stress at each measurement location was performed using a Monte Carlo model. Local velocity uncertainty varied greatly and depended largely on local conditions such as particle seeding, velocity gradients, and particle displacements. Uncertainty in low velocity regions in the sudden expansion section of the nozzle was greatly reduced by over an order of magnitude when dynamic range enhancement was applied. Wall shear stress uncertainty was dominated by uncertainty contributions from velocity estimations, which were shown to account for 90 - 99% of the total uncertainty. This study provides advancements over the previous work through increased PIV image resolution, improvements in velocity measurement accuracy, calculations of wall shear stress, and uncertainty analyses for both velocity and wall shear stress measurements. The velocity and shear stress data, with appropriate uncertainty estimates, will be useful for performing comprehensive validation of current and future CFD simulations of the nozzle model.

# Introduction

Computational fluid dynamics (CFD) is potentially a valuable tool for the validation of safety and efficacy levels in medical devices. Specifically, CFD can deliver a fast and low-cost alternative to animal studies during the initial stages of product development, as well as provide supplemental data to help substantiate safety claims during Food and Drug Administration (FDA) product reviews. However, CFD is not considered a stand-alone method for device safety validation due to the method's large dependence on boundary conditions and modeling assumptions.

In-vitro techniques, such as Particle Image Velocimetry (PIV), can complement CFD by providing experimentally derived velocity estimates within bench-top models that recreate the medical device and physiological conditions. PIV is especially attractive for medical device validation as the method is non-invasive and can provide high spatial and temporal resolution. Furthermore, many of the hydrodynamic quantities relevant to medical device safety, including wall shear stresses (WSS), viscous shear stresses (VSS) and Reynolds stresses, may be estimated from velocity gradients and fluctuations calculated from PIV velocity fields.

Many studies have demonstrated the relationship between abnormal shear in medical device implants and potential adverse patient events such as hemolysis, platelet activation, and thrombus formation [1-6]. The exact shear level at which hemolytic blood damage occurs depends upon exposure time, due to the viscoelastic behavior of red blood cells (RBC) [1, 7, 8]. However, Baldwin et al. predicted that a threshold of 1500-4000 dynes/cm$^2$ [9] marks the initiation of RBC damage, and other studies have shown that platelet activation occurs below these values [1]. This range is often surpassed in commonly used medical devices and vascular anomalies, including mechanical heart valves (MHVs) [1-5] (used alone and in artificial hearts), and in stenotic arteries [8, 10]. Specifically, artificial hearts are associated with abnormally high shear regions (on the order of 10,000 dynes/cm$^2$) due to regurgitating jets impinging on closed MHV leaflets [1], but can also cause problems due to low WSS (<15-20dyne/cm$^2$ [11, 12]) in the chambers of the device. Pro-thrombogenic flow properties such as stagnation regions, high fluid residence times and diminished WSS can occur in bifurcating vessels, cardiovascular stents, and anastomoses from both bypass grafts and Fontan procedures. Such environments can result in atherosclerotic lesions, intimal hyperplasia, in-stent restenosis, and thromboemboli [13-19]. These examples support the need for robust and accurate medical device validation techniques that combine experimental and numerical efforts.

Unfortunately, the measurement of shear and velocity fields in medical devices using PIV can be challenging due to large velocity gradients at boundaries, high dynamic ranges of velocity, and low signal-to-noise ratios from poor image quality. These complications test the limits of PIV by increasing measurement uncertainty, thus calling into question the ability of conventional PIV to provide a validation dataset for CFD.

The aim of this study is to establish a protocol for producing velocity and WSS measurements with uncertainty estimates for CFD validation using advanced PIV processing and post-processing techniques. The work is an extension of a previous study [20] using the FDA benchmark nozzle model (Fig. 1); however, several experimental advancements that extend beyond the scope of the previous study were utilized to establish a more comprehensive CFD validation dataset.

First, temporally resolved PIV data was acquired to allow for dynamic range enhancement processing to improve measurement precision in low velocity regions such as near boundaries and in recirculation regions [21]. Additionally, ensemble and phase correlations were used to augment correlation signal-to-noise ratios (SNRs), thus allowing for a reduction in interrogation window size, increased spatial resolution, and reduced near wall bias in comparison to the previous study [22-25]. The accuracy enhancements from the aforementioned processing techniques improved measurement reliability of WSS, which was unavailable in the former work. Radial basis functions (RBF) were used to compute gradients for VSS and WSS estimations due to the method's low susceptibility to noise and high accuracy in comparison to finite difference schemes. Additionally, the absence of reported uncertainty for the previous study limited comparisons with CFD predictions, whereas the current work provides a means of spatial comparisons for both velocity and WSS estimations. Finally, this work presents an analysis of Reynolds stresses and energy dissipation rate as well, and provides estimates of pressure fields derived from PIV data for order of magnitude comparisons with CFD. Many of the advancements performed in this study rely on PIV processing and post-processing techniques, and therefore, a brief description of each method will be discussed.

Conventional PIV image processing utilizes discrete window offset (DWO) by shifting image windows by a distance equal to the integer-pixel displacement [26]. The use of image deformation, which continuously deforms images based on velocity information computed from previous processing passes, has become widely used and is considered a standard by many PIV users [27]. This technique is especially advantageous in high shear regions, such as near walls, where images are deformed to account for variations in particle displacements within the same interrogation window. Additionally, various windowing schemes such as apodization filters remove wraparound aliasing caused by the periodicity assumption when performing fast Fourier transforms [28]. In addition to these conventional processing techniques, which were used in the former work [20], a number of recent advancements, including phase filters, dynamic range enhancement, and ensemble correlation, were applied to the current dataset to improve the accuracy of velocity and velocity gradient estimations and to determine measurement uncertainty bounds.

Phase filters, such as the phase-transform (PHAT) filter, can be implemented to increase the correlation signal-to-noise ratio (SNR) and are especially valuable in data sets with high background noise levels, inhomogeneous illumination, and image artifacts [29, 30]. Many of the aforementioned obstacles are often encountered in medical device models, including the analog device used in this study. Specifically, because all information regarding particle displacement is contained in the phase space, correlations are performed in the spectral domain to exclude noise embedded in the magnitude information of conventional cross-correlations. The resulting correlations have highly defined peaks, similar to the Dirac delta function, as well as increased SNRs. The Robust Phase Correlation (RPC), which was also used in this study, takes the analysis one step further by analytically recreating the PIV SNR by considering sources of signal and noise to improve sub-pixel accuracy [31, 32].

Secondly, dynamic range enhancement, or multi-frame PIV, utilizes the flexibility of time-resolved PIV to improve cross-correlation accuracy in data sets with large velocity ranges. This method was applied to the current work to reduce measurement uncertainty near boundaries and within the recirculation region of the sudden expansion where sub-pixel particle displacements and large bulk flow velocities occur in the same domain. Specifically, lower velocity regions in the flow field are computed using longer temporal separations than in high-speed regions. The method improves measurement precision in low velocity regions by increasing the particle displacement above sub-pixel levels [21]. The frame separation used at a given vector location is determined by the criteria

$$Q = \frac{c_1}{c_2} \times \left[ \frac{\text{PIV error}}{\text{velocity magnitude}} \right], \quad (1)$$

where Q is the quality for a given image pair, $c_1$ and $c_2$ are the primary and secondary peaks of the correlation at each vector location, PIV error is the user defined allowable error and velocity magnitude is the pixel displacement between the two chosen frames. The temporal separation resulting in the maximum quality is chosen for the final velocity field.

Finally, ensemble correlation performs temporal averaging of the PIV cross-correlations at each measurement location prior to sub-pixel interpolation to increase the correlation SNR [33]. This method has been widely used for velocity estimations in steady flow data sets and is particularly advantageous in regions with low seeding densities or poor image quality where the signal is diminished, as seen at near wall locations and in the recirculation region of the current model. In the current study, ensemble correlation allowed for an increase in spatial resolution in comparison to the previous study [20] as quantified in the following section. Furthermore, by directly averaging correlations as opposed to velocity measurements, bias from erroneous instantaneous measurements can be reduced.

It is clear that recent advances in PIV analysis have led to improvements in measurement accuracy; however, only a handful of recent studies focus on the quantification of uncertainty [34-36]. Timmins et al. discussed the existence of several spatially and temporally varying factors that contribute to the non-uniformity of PIV error, including the PIV algorithm, hydrodynamic environment, and experimental setup. Their work suggests that a universal estimate for PIV error is inadequate to describe an entire flow environment, and they present a method for estimating uncertainty at each vector location resulting from several sources.

Additionally, Sciacchitano et al. developed an *a posteriori* technique, which uses the residual distance between matched particles in image pairs (after cross-correlations) to estimate spatially distributed uncertainty [35]. Finally, Charonko et al. utilized the PIV cross-correlation to identify a relationship between measurement uncertainty and correlation peak ratio,

$$\delta \Delta x = \left( 9.757 e^{-\frac{1}{2}\left(\frac{P-1}{1.139}\right)^2} \right) + \left(1.405 P^{-1}\right)^2 + \left(1.72 \times 10^{-5}\right)^2, \quad (2)$$

where $\delta \Delta x$ is the 95% uncertainty bound for RPC correlations (in pixels) and P is the ratio of the first to second highest peak of the correlation. This technique provides an uncertainty estimate for the cross-correlation at each vector location in time to account for the non-uniformities discussed above.

Ultimately, velocity measurements with associated uncertainties will be used to determine near wall gradients for WSS estimates. Several options exist for gradient estimators such as high-order finite differencing and analytical fitting methods [37-40]. Finite differencing methods provide a straightforward technique to determine gradients between gridded vector locations. However, analytical fitting methods such as Radial Basis Functions (RBF) can result in a reduction in error and susceptibility to noise in comparison to finite difference approaches. The method uses local data points to approximate the data surface as a linear combination of RBFs and polynomial bases. An analytical expression of the data surface is determined from the base coefficients and ultimately differentiated to compute local gradients.

## Materials and Methods

### Experimental Methodology

Particle image velocimetry was performed on the FDA benchmark nozzle model, which has been thoroughly described in previous literature [20, 41] and is illustrated in Fig. 1(a). The geometry chosen for analysis contains several features that commonly exist in fluidic medical devices such as sudden and gradual (conical) expansions and contractions. The work herein is limited to the flow direction producing a conical contraction and sudden expansion depicted in the figure as left-to-right flow. For consistency, all axial and radial locations are reported with respect to the centerline at the sudden expansion as indicated in Fig. 1(a), with positive $x/D_{th}$ locations occurring downstream of the sudden expansion, where $D_{th}$ represents the throat diameter, 4 mm.

Three sides of the model were polished to obtain optical access to the channel while the top side contained counter-bored pressure ports (0.56mm in diameter) centered radially along the axis of the nozzle. Slight optical distortions were present at junctions between the inlet, contraction, throat and sudden expansion, but were most notable at the transition between the throat and sudden expansion. Optical distortions from the wall curvature were largely eliminated through refractive index matching using a sodium iodide solution for the working fluid. The blood analog was composed of approximately 50 wt % saturated Sodium Iodide - DI water solution mixed with a 60-40 wt % mixture of glycerin and water. The refractive index of the solution was measured using a refractometer (Misco Palm Abbe-202) and adjustments were made to the solution throughout the experiment to maintain a refractive index between 1.485-1.486. A working fluid density of 1700 kg/m$^3$ and viscosity of 3.13 cP were measured using a digital scale (OHaus Scout II, Parsippany, NY) and viscometer (Brookfield LVDC-II+P, Middleboro, MA).

Steady flow was created using gear pumps to reach the desired flowrate and located upstream from a flow straightener (Fig. 2). Stainless steel tubing (12mm ID) provided entrance and exit lengths of approximately 150D and 60D respectively, where D is equal to the inlet nozzle diameter of 12mm. Pressure data was acquired at ten pressure tap locations (Fig. 1 (b)) using differential pressure transducers (0-2 psi, Setra 210, Boxborough, MA) throughout the duration of each test case. Flow rate was also measured throughout the duration of each test case using an ultrasonic flow meter located downstream of the test section (Transonic Systems T110, probes 11XL and 3XL). The RMS of the measured flow rate showed negligible variations in the flow rate (<4mL/min) for all cases (range of flow rates was 415 to 6630 mL/Min). Four Reynolds numbers, based on the throat diameter ($D_{th}$) and measured fluid properties, were chosen for analysis ($Re_{th}$ = 500, 2000, 5000, and 8000).

PIV was performed at five locations in the nozzle as illustrated in Fig. 1(b). An intensified IDT Xs-5i high-speed camera and a high-speed Nd:YAG laser (Dantec Lee, $\lambda$=532 nm) were used to acquire time-resolved PIV data at a frame-pair rate of 250 Hz with 7 μm polystyrene particles. The resulting spatial resolution was 15.1 μm/px and the frame separation varied for each Reynolds number to obtain a maximum displacement in the jet region of approximately 20 pixels between image pairs.

### **PIV Processing Methodology**

Prior to PIV processing, images underwent a temporal minimum intensity subtraction at each pixel location to remove background illumination. PIV processing was performed using Prana (available at 'http://sourceforge.net/projects/qi-tools/'), an in-house, open-source software program which has been thoroughly validated in several publications [34, 42-44].

First, processing was conducted using ensemble and multi-frame correlation techniques to resolve the large dynamic range in the flow fields. As described above, multi-frame correlation utilizes multiple frame separations to resolve high and low velocities. The frame separations chosen for the current work varied for each case and depended on the frame-pair and inter-pair sampling rates. This process resulted in displacements between 1-20 pixels throughout each flow field. Two-pass processing was performed using the RPC on a total of 4300 images with window deformation [27, 32]. The window resolution of the first pass was 64 x 16 pixels with a grid resolution of 16 x 8 pixels where interrogation windows were extended in the primary direction of flow. The second pass used a window resolution of 16 x 16 pixels and a grid resolution of 4 x 4 pixels. The previous benchmarking study [20] resulted in a final physical grid resolution ranging between 110-220 μm/vector; through the use of ensemble correlation, the current study was able to achieve a resolution of 60 μm/vector, approximately a 2-fold increase. To ensure that a proper deformation was applied to subsequent passes, two iterations were performed on the first pass and three on the second. For all passes, a Gaussian apodization filter was applied to the windows to reduce wrap-around aliasing [28]. PIV validation was performed using the Universal Outlier Detection (UOD) scheme [45] with neighborhoods of 7x3 and 3x3 vectors for the first and second passes, respectively.

Secondly, time-accurate processing was conducted to compute Reynolds stresses, energy dissipation rate, and pressure fields from PIV data. Instantaneous processing was executed using multi-frame RPC correlations with window deformation and Gaussian apodization windows as applied for the ensemble correlation processing above. However, to compensate for the reduction in SNR with respect to ensemble correlation, interrogation windows were enlarged to 96 x 32 pixels with 32 x 16 pixel grid resolution on the first pass, and 32 x 32 pixel windows with 16 x 8 grid resolution on the second pass. As in the ensemble case, two iterations were performed on the first pass and three iterations on the second, and all outlier detection was performed using the UOD scheme.

## Wall Shear Stress Estimation

After PIV processing was completed, wall shear stress was estimated from the velocity fields by determining the near wall velocity gradients. First, particle images were summed in time and wall detection was performed by locating channel boundaries from the edges of the illuminated regions of the summation image. Resulting pixel wall locations were converted to fractional vector locations prior to velocity gradient estimations. Wall shear stress is defined as

$$\tau_w = \mu \dot{\varepsilon}'_{12} \tag{3}$$

where $\mu$ is the dynamic viscosity and $\dot{\varepsilon}'_{12}$ is the wall-tangential components of the strain rate tensor. Specifically,

$$\dot{\varepsilon}'_{12} = \left[ \frac{\partial u}{\partial y} + \frac{\partial v}{\partial x} \right] \tag{4}$$

where $u$ and $v$ are the streamwise and spanwise velocities, respectively. A rotation transformation matrix,

$$T_{ij} = \begin{bmatrix} \cos\theta & \sin\theta \\ -\sin\theta & \cos\theta \end{bmatrix} \tag{5}$$

was applied to the strain rate tensor to account for any misalignment between the image coordinate system and the coordinate system along the wall, where $\theta$ is the angle between the wall and the image coordinate systems. Here, the transformation is applied as $\dot{\varepsilon}'_{mn} = T_{mi}T_{nj}\dot{\varepsilon}_{ij}$ as described by Charonko et al. [12, 25].

The velocity gradients used in equation (4) and in VSS calculations were computed using a Thin Plate Spline (TPS) Radial Basis Function (RBF). A TPS was chosen over other options of RBF's as this function provides the smoothest possible interpolated surface for a given set of data points [37]. Velocity gradients at each wall location were interpolated from a 7x7 grid of data points that were centered at each vector query point but extended only inward from the wall into the flow field. Erroneous velocity measurements as determined by the UOD and measurements with peak ratios less than 1.5 were excluded from the RBF during WSS calculation. WSS was not reported for query points with erroneous measurements or insufficient peak ratios ($c_1/c_2 < 1.5$).

## Velocity and Shear Stress Uncertainty Analysis

Uncertainty bounds were estimated for velocity measurements using Taylor series expansion, giving

$$\partial v = \sqrt{\left(\frac{\partial v}{\partial M}\delta M\right)^2 + \left(\frac{\partial v}{\partial t}\delta t\right)^2 + \left(\frac{\partial v}{\partial \Delta x}\delta \Delta x\right)^2} \quad (6)$$

where $\partial v$ is the uncertainty of the velocity, $v$, and includes contributions from the PIV cross-correlation, $\delta \Delta x$ (from Eq.2), image magnification, $\delta M$, and acquisition timing $\delta t$. For ensemble cases, the uncertainty of the velocity due to the PIV cross-correlation ($\delta \Delta x$) was obtained from the resulting ensembled correlation plane. In contrast, for the time averaged data, $\delta \Delta x$ was estimated for each image pair to compute a mean uncertainty due to the PIV cross-correlation. In addition, the uncertainty contribution from image magnification was assumed to be constant across the PIV images and was determined using calibration images taken of the FDA model nozzle. Here, a measurement uncertainty in the calibration image was assumed to be ±1pix, the resolution of the camera, and the uncertainty of the physical model dimension was

±1/100", based on fabrication tolerances. Specifications from the laser-camera timing system were used to compute the uncertainty from acquisition timing (±2.1x10$^{-8}$ s).

Monte Carlo simulations [46] were performed in Matlab to determine uncertainty on WSS measurements due to the challenge of determining analytical derivatives of WSS with respect to uncertainty sources in the velocity gradient estimation. Specifically, this task is difficult because velocity gradients are estimated from a neighborhood of velocity data points using a TPS, thus complicating the relationship between uncertainty and WSS. In performing Monte Carlo simulations, distributions of 10,000 data points for image magnification, acquisition timing, cross-correlation shifts, wall locations, and viscosity were generated and used as input parameters for the RBF, which was iterated 10,000 times. Uniform distributions were used for calibration image measurements and the acquisition timing with bounds equal to the uncertainty values described above for these parameters. Uniform distributions were also chosen to represent uncertainty in wall detection and viscosity with bounds equal to ±1pix, again based on camera resolution, and 1.1x10$^{-5}$ cP (1% full scale dependent on the operating speed of the viscometer, in this case 60 RPM), respectively. The average values of the above distributions were equal to the values measured experimentally. Distributions for particle image displacements were limited to the direction estimated by the local ensemble velocity with the mean value equal to the measured velocity and two standard deviations equal to the uncertainty determined in Eq. 2. Resulting distributions of particle displacements were non-Gaussian; however, the final WSS distributions were normal. The five and ninety-five percentiles were determined based on the resulting WSS distributions and reported as the 95% confidence intervals.

## Results

Resulting velocity profiles at the inlet (x = -21D$_{th}$), normalized by the average inlet velocity, are plotted in Fig. 3 (a & b) for ensemble correlation and time-averaged fields. Uncertainty bars indicate 95% confidence intervals as computed from equations (2) and (6). For the time-averaged field, uncertainty from each instantaneous field was averaged in time to compute the final uncertainty value. Profile symmetry was determined using a symmetry index (SI) computed as the ratio of the flow rates in the left and right halves of each profile [20]. Symmetry indices indicated symmetric flow with SI values greater than 0.9 for ensemble and instantaneous profiles (values are listed in Table 1 with corresponding local inlet pipe Reynolds numbers listed in parentheses). Profiles for throat Reynolds numbers less than 5,000 (local inlet Re<1,670) resemble Poiseuille flow while the highest throat Reynolds number case ($Re_{th}=8000$) resembles plug flow, with a local inlet Reynolds number of 2,670. In general, the ensemble correlation and time-averaged profiles are in agreement. However, a larger near-wall bias is observed for the time-averaged profiles in comparison to the ensemble correlation profiles (inset plots of Fig. 3 (a & b)), as well as increased uncertainty throughout the time-averaged profiles.

Velocity profiles normalized by the average throat velocity are plotted in Fig. 4(a & b) at locations $x = 2D_{th}$ and $x = 5D_{th}$ in the downstream expansion region after ensemble correlation processing. When comparing results of the two locations, profiles at $x = 5D_{th}$ appear to widen in comparison to the upstream location where shear layer gradients are diminished and a more prominent recirculation region is observed.

Centerline velocities normalized by the inlet average velocity are plotted along the length of the nozzle for all Reynolds numbers and shown in Fig. 5. Optical distortions created from the nozzle fabrication process prevented reliable PIV measurements at the entrances to the throat and sudden expansion, and therefore centerline velocities are not reported at these locations. Normalized centerline velocities increase from a value of approximately 2 at the inlet to a maximum value at the sudden expansion ($x/D_{th} = 0$), and then reduce throughout the remainder of the measured downstream area.

The velocity uncertainty fields for Re = 500 at the inlet and sudden expansion regions are shown in Fig. 6(a & c). The inlet field of view is cropped due to visibility problems upstream from $x = -22D_{th}$. In Fig. 6, unreliable data, as determined by the UOD, and insufficient peak ratios ($c_1/c_2<1.5$) are excluded and contoured in white. For both fields, uncertainty is increased at near wall locations, and is particularly observable near the upper and lower corners of the sudden expansion where particle seeding was low and unreliable velocity data was obtained. Additionally, increased uncertainty is observed downstream of the sudden expansion in the high shear region surrounding the jet, which decreases with increasing $x/D_{th}$. For brevity, the 5, 50, and 95 percentiles for velocity uncertainty normalized by the average inlet velocity are listed in Table for all cases with dynamic range enhancement. Here, the average inlet velocity was calculated from the measured flow rate.

To illustrate the influence of dynamic range enhancement processing, uncertainty fields without multi-frame processing are shown in Fig. 6 (b & d). Here, velocities are computed conventionally, with only PIV image pairs separated by the shortest time-frame separation. When comparing uncertainty between conventional and multi-frame processing, a significant reduction in uncertainty is observed for low velocity measurements, particularly in the recirculation regions of Fig. 6(c & d) and near-wall locations in the inlet (Fig. 6(a & b)) where uncertainty is observed to reduce by over an order of magnitude in some regions.

Profiles of VSS normalized by kinetic energy, $1/2\, \rho\, v_{th}^2$, are plotted in Fig. 7(a & b) at downstream locations $x = 2D_{th}$ and $x = 5D_{th}$ where $v_{th}$ is the average throat velocity based on the measured flow rate. As expected, profiles appear relatively symmetric and opposite in sign across the centerline of the flow field. Peak VSS decreased as the flow progressed downstream of the sudden expansion for all Reynolds numbers as depicted in Fig. 7(c).

WSS and corresponding 95% uncertainty bounds were calculated from the ensemble correlation field and are reported in Fig. 8 (a & b) along the model walls in the gradual contraction and throat regions. General trends of the WSS in these regions agree with expected results showing increasing WSS for increasing Reynolds number. The gradual contraction indicates a rapid increase in WSS as the nozzle diameter reduces. WSS values in the throat are seen to gradually decrease with increasing $x/D_{th}$ for all Reynolds numbers as velocity profiles begin to develop and near wall gradients reduce. For both cases, uncertainty bounds are observed to fluctuate along the length of the wall, and occasional spikes in uncertainty are visible due to small fluctuations in measured velocity (prominent in Fig. 8 (b)).

The median uncertainties with respect to locally computed WSS values are reported in Table 3 for all Reynolds numbers and PIV locations. Uncertainty computed without contributions from the velocity estimation are listed in parentheses for comparison. Large uncertainties, are reported for the inlet, gradual contraction, and sudden expansion regions, while lower uncertainties are reported for the throat regions. No obvious trends are observed in terms of percent uncertainty with respect to the various Reynolds numbers, although a clear increase in uncertainty bound values is seen for higher Reynolds number cases in the aforementioned line plots. Bounds estimated by neglecting velocity uncertainty indicate a marked decrease in the estimated bounds, as uncertainty from velocity estimations accounted for approximately 90 – 99% of the total WSS uncertainty.

Pressure measurements acquired for each Reynolds number were averaged in time and across the five data sets taken for each flow case and are plotted in Fig. 9(a). Fields of relative pressure were estimated from instantaneous PIV velocity data using an omni-directional line integral method outlined by Charonko et al. and adapted from Liu and Katz [47, 48]. The pressure fields in the gradual contraction and throat are shown in Fig. 9(b & c) for Re = 8000. The fields are referenced with respect to the measured pressure at $x/D_{th}$ = -11.5 and $x/D_{th}$ = -5 for the contraction and throat, respectively. As expected, pressure is seen to decrease as the velocity increases in the throat. Additionally, centerline pressures from pressure field predictions are also plotted in Fig. 9 (a) showing generally good agreement with measured pressures.

Reynolds shear stress profiles for $\langle u'^2 \rangle / U_o^2$ and $\langle v'^2 \rangle / U_o^2$ derived from instantaneous PIV fields are plotted in Fig. 10 (a & b) at several $x/D_{th}$ locations for Reynolds numbers 5,000 and 8,000. Here $U_o$ is defined as the centerline throat velocity. For both of the presented cases, velocity fluctuations in the streamwise direction are dominant, and stress profiles are observed to widen with a reduction in peak values for increasing values of $x/D_{th}$. Profiles of normalized energy dissipation rate ($\varepsilon D_{th} / U_o^3$) shown in Fig. 10(c), indicate high rates of dissipation localized to the shear layer, which spread with increasing $x/D_{th}$. Additionally, the peak rate of energy dissipation appears to grow with $x/D_{th}$ for $Re_{th}$ = 5000 while decreasing slightly for $Re_{th}$=8000. Dissipation was calculated using a two dimensional representation assuming fully resolved, turbulent, isotropic, and homogeneous fields [49, 50]. Although the minimum resolution for this data (~240μm) is approximately 10 times the estimated Kolmogorov length scale (~25μm), and the flow environment may not be considered isotropic or homogeneous, the analysis still provides an order of magnitude estimate of dissipation and dissipation patterns.

## **Discussion**

This study integrated several advanced PIV processing techniques that can improve measurement accuracy for velocity and shear estimations, including ensemble correlation, dynamic range enhancement, phase correlations, and RBFs. Many of these techniques are particularly useful in flow environments with large velocity gradients, a high range of velocity magnitudes, and poor image quality; conditions that are commonplace in many blood-contacting medical device studies. This study also provided a methodology for estimating uncertainties in viscous shear stress measurements and implements a velocity uncertainty estimation method developed by Charonko et al. [34]. With increasing use of CFD in medical device design and regulatory submissions, there is a growing need for developing standardized methodologies to verify and validate (V & V) CFD models. A critical but often ignored stage in the V & V process is the estimation of experimental uncertainties [51]. The advantage of this uncertainty estimator [34] is that it can provide the uncertainty values in a relatively fast and uncomplicated manner. Specifically, the displacement uncertainty due to the PIV correlation was obtained using a single equation using the correlation peak ratio, and the extension to wall shear stress and viscous shear stress was performed straightforwardly using a Monte Carlo Simulation. Consequently, this uncertainty estimation method will help in performing a more comprehensive validation of the CFD results.

One of the prominent challenges in the current flow model is the presence of large velocity gradients at near-wall locations in the inlet, conical contraction, and throat regions. Inlet velocity profiles for ensemble correlation data in Fig. 3 (a) demonstrate a reduction in uncertainty and near-wall bias in comparison to time-averaged data shown in Fig. 3 (b). The improvements in uncertainty occur due to strengthened correlation SNRs produced by correlation summation, which allows for diminished window sizes and an increased spatial resolution of approximately 3 times larger than the previous study [20]. Conversely, time-averaged fields computed from instantaneous processing schemes require larger windows to combat low SNRs from sparse seeding near the wall. The implication of increased window size and lower resolution is two-fold. First, large interrogation windows increase the range of particle displacements in shear flow, widening the cross-correlation [52], and in extreme cases can cause peak-splitting [53], both of which lead to increased uncertainty. Secondly, because particle displacements are averaged across interrogation windows, a larger velocity bias is present at the wall as seen in the insets of Fig. 3(a & b). Ultimately, if CFD validation is to take place at near-wall locations, the improvements gained through ensemble processing should be employed whenever possible.

Poor image quality, sparse seeding, and the overlap of interrogation windows with regions outside of the flow environment, also provide challenges at boundaries by introducing noise and reducing the available signal. In cases such as these, phase correlations can be implemented to reduce the influence of noise by performing correlations in the spectral domain. Although a direct comparison with standard correlation methods is not provided in this work, other works have sufficiently shown the reduction in correlation noise floor, higher peak detectability, and increased accuracy resulting from this procedure [31, 32, 42]. However, despite the use of phase and ensemble correlations with deformable windows, near-wall velocity estimates continue to be associated with higher uncertainty in comparison to the bulk flow, as shown in Fig. 6 (a).

Steep velocity gradients are also observed away from boundaries such as in the jet shear layer shown in velocity profiles (Fig. 4a) and VSS profiles (Fig. 7(a)) plotted at $x = 2D_{th}$. Downstream at $x = 5D_{th}$, the jet dispersion is reflected in velocity (Fig. 4b) and VSS (Fig. 7b) profiles, which widen in comparison to $x=2D_{th}$ and display diminished peak velocity, velocity gradients, and VSS. The corresponding velocity uncertainty (Fig. 6(c)) for Re = 500 resembles patterns of VSS with increased uncertainty occurring in the shear layer, and a slow decline in uncertainty following the decay of maximum VSS with increasing $x/D_{th}$ (Fig. 7(c). Particularly, this reflection of uncertainty trends on VSS patterns isolates the influence of shear flow on velocity uncertainty without influences from boundary complications.

The sudden expansion location is complex due to the high range of velocities between the jet and recirculation zones. For this data, a dynamic range of approximately 100:1 was observed between the jet and recirculation regions, resulting in displacements on the order of 0.1pixels in the low flow regime. Through the use of dynamic range enhancement, a larger frame separation lengthened particle displacements above 1 pixel, allowing for higher precision velocity estimations. In extreme cases, such as areas of the sudden expansion, this correction resulted in a reduction of local uncertainty from over 1000% to values below 5% of the local velocity, thereby providing more reliable measurements.

WSS measurements along the throat walls in Fig. 8(b) are associated with lower uncertainty in comparison to other regions. This result is somewhat unexpected due to the high velocity gradients in this region. However, adequate seeding, coupled with phase correlations, small and deformable interrogation windows, and RBF derivative schemes, most likely contribute to the improved performance in this region. Nonetheless, isolated spikes in uncertainty coincide with small WSS fluctuations, again highlighting the sensitivity of WSS to even moderate levels of velocity uncertainty.

The overwhelming influence of velocity uncertainty on WSS measurements is apparent when comparing uncertainty bound estimates with and without contributions from velocity estimates (Table 3). In part, this sensitivity is not surprising when considering the exponential propagation of uncertainty through spatial or temporal derivatives from PIV fields even with minimal error [37, 38, 47]. As expected, the challenges discussed above, including sparse seeding, high velocity gradients and poor image quality, largely influence WSS measurement uncertainty. However, the results are surprising when considering the small contributions from velocity gradient estimations. These findings reiterate that, in general, the RBF provides a robust method to calculate gradients with moderate velocity fluctuations, here introduced in Monte Carlo simulations as local uncertainty. Furthermore, although the use of RBFs is seemingly overcomplicated in comparison to central difference schemes, the procedure is similar to that of finite difference methods, which also require interpolation of the gradients at non-gridded wall locations, as well as coordinate transforms to properly account for wall-tangential velocity components. This argument can be extended to WSS uncertainty analysis where both methods require Monte Carlo type analyses as performed in this work.

In addition to WSS, Reynolds stresses remain an important metric to evaluate blood-contacting medical devices due to potential RBC damage caused by shearing forces. For the current geometry, velocity fluctuations are observed to be largest in the jet shear layer downstream of the sudden expansion. Specifically, Reynolds stress profiles for Re = 5000 and 8000 in Fig. 10 a & b indicate dominant velocity fluctuations in streamwise rather than spanwise components. As expected, peak fluctuations occur in the shear layers and profiles widen as the jet disperses for both Reynolds numbers. For Re=8000, peak fluctuations remained relatively constant across the domain. Additionally, profiles of the rate of energy dissipation for this Reynolds number, shown in Fig. 10 (c), widen with increasing $x/D_{th}$. For Re = 5000, peak fluctuations and energy dissipation appear to grow throughout the measured region with increasing $x/D_{th}$. Despite the assumptions made regarding measurement resolution and isotropic, homogeneous flow conditions, this analysis permits an order of magnitude comparison for CFD validation. Supporting this assumption, work by Sharp and Adrian found, using the same dissipation model, that PIV with spatial resolution seven times greater than the Kolmogorov scale was consistently capable of capturing at least 70% of the true dissipation [50].

Finally, estimates of relative pressure fields are included in this analysis to provide an additional reference for CFD comparisons. General agreement was observed among relative pressures measured at port locations and that of the estimated fields. However, sensitivity to velocity uncertainty was reflected in pressure estimates, which displayed increased deviations from expected values near regions of high velocity uncertainty. Nonetheless, the reported pressures should supplement the PIV comparison with CFD.

## **Conclusion**

The current work draws attention to the challenges of gathering high quality PIV data. Although the geometry and flow conditions are simplified when compared to many real-world medical device applications, the implemented techniques are easily extended into more complex and time-varying methodologies. Nonetheless, the specific flow conditions and model geometry demonstrate measurement challenges that occur near boundaries due to increased optical distortions, high velocity gradients, and large ranges of velocity. For CFD validation purposes, such locations are often areas of interest to estimate boundary conditions for use in simulations, and therefore, data quality in these regions is critical.

This study synergistically implements several advanced PIV processing and post-processing techniques including ensemble correlation, dynamic range enhancement, phase correlations, and RBFs to improve velocity and shear estimations for validation of CFD simulations of flow in medical device models.  In particular, the analysis focused on flow fields with large near-wall velocity gradients, high ranges in velocity, and poor correlation SNRs resulting from sparse particle seeding and/or poor image quality, which are problems in many medical device studies using PIV.  An uncertainty analysis was provided in a spatial sense to more comprehensively compare velocity and WSS PIV estimates with CFD predictions.  Results indicate that WSS uncertainty is primarily dominated by velocity uncertainties, highlighting the importance of advanced PIV processing schemes to improve accuracy at boundary locations.  Of the many challenges inherent to medical device validation, sparse near-wall seeding is perhaps the most difficult to resolve due to lack of signal.  Alternatively, the use of advanced processing and post-processing schemes can often mitigate problems such as large velocity ranges and gradients, as well as noisy images.  This study supplements previous works by addressing protocols for measurement of velocity, WSS, uncertainty, Reynolds stresses and dissipation and pressure estimates to provide a comprehensive dataset for CFD validation.

## **Acknowledgements**

This work was supported by the National Science Foundation Scholar-in-Residence program at the Food and Drug Administration, NSF-SIR award number 1239265.  Additionally, the authors would like to acknowledge contributions from Dr. Steven Day (Rochester Institute of Technology), Matthew Giarra (Virginia Tech), and Dr. Sandy Stewart (FDA).

# Nomenclature

| | |
|---|---|
| BMS | bare metal stents |
| CFD | computational fluid dynamics |
| DES | drug eluting stents |
| $D_{in}$ | Inlet diameter, 12mm |
| $D_{th}$ | Throat diameter, 4mm |
| DWO | Discrete Window Offset |
| ISR | in-stent restenosis |
| PHAT | Phase-only filter |
| PIV | particle image velocimetry |
| RBC | red blood cell |
| $Re_{in}$ | Reynolds number based on $D_{in}$ |
| $Re_{th}$ | Reynolds number based on $D_{th}$ |
| RPC | Robust Phase Correlation |
| SNR | Signal to Noise ratio |
| $v_{in}$ | average inlet velocity |
| $v_{th}$ | average throat velocity |
| VSS | viscous shear stress |
| WSS | wall shear stress |

# Tables and Figures

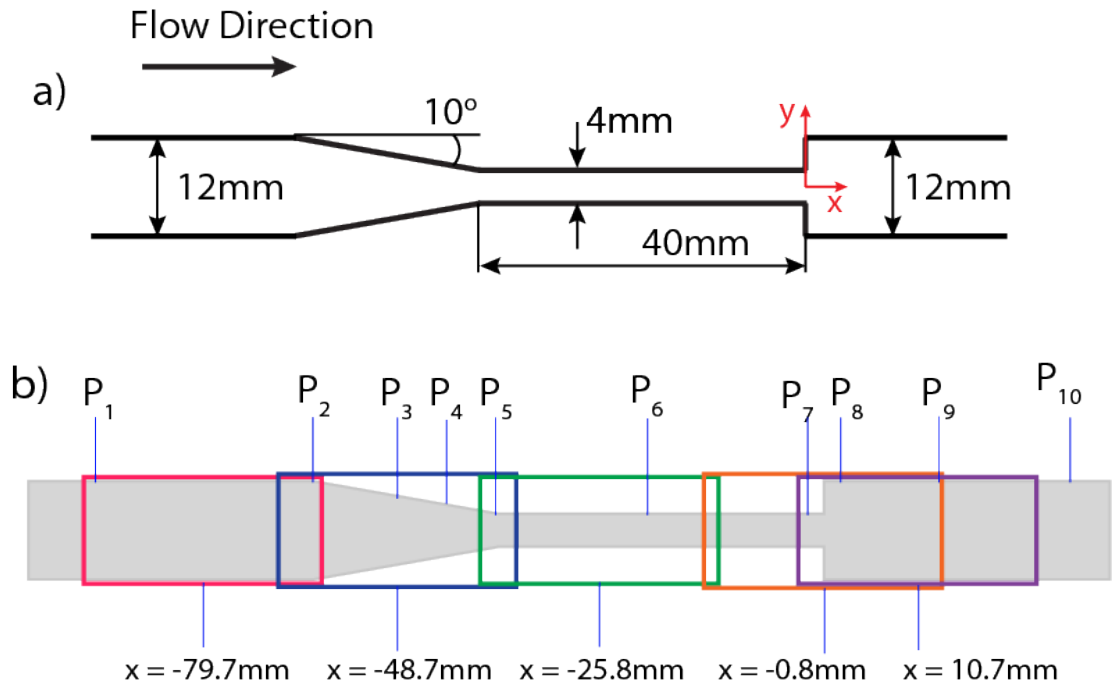

**Fig. 1: (a) Dimensions of FDA benchmark nozzle model, flow direction left-to-right. (b) Locations of pressure (P) measurement taps and five overlapping field of view locations acquired in PIV analysis. The center of each field of view is labeled below the diagram and each field of view is 12.6mm x 35mm.**

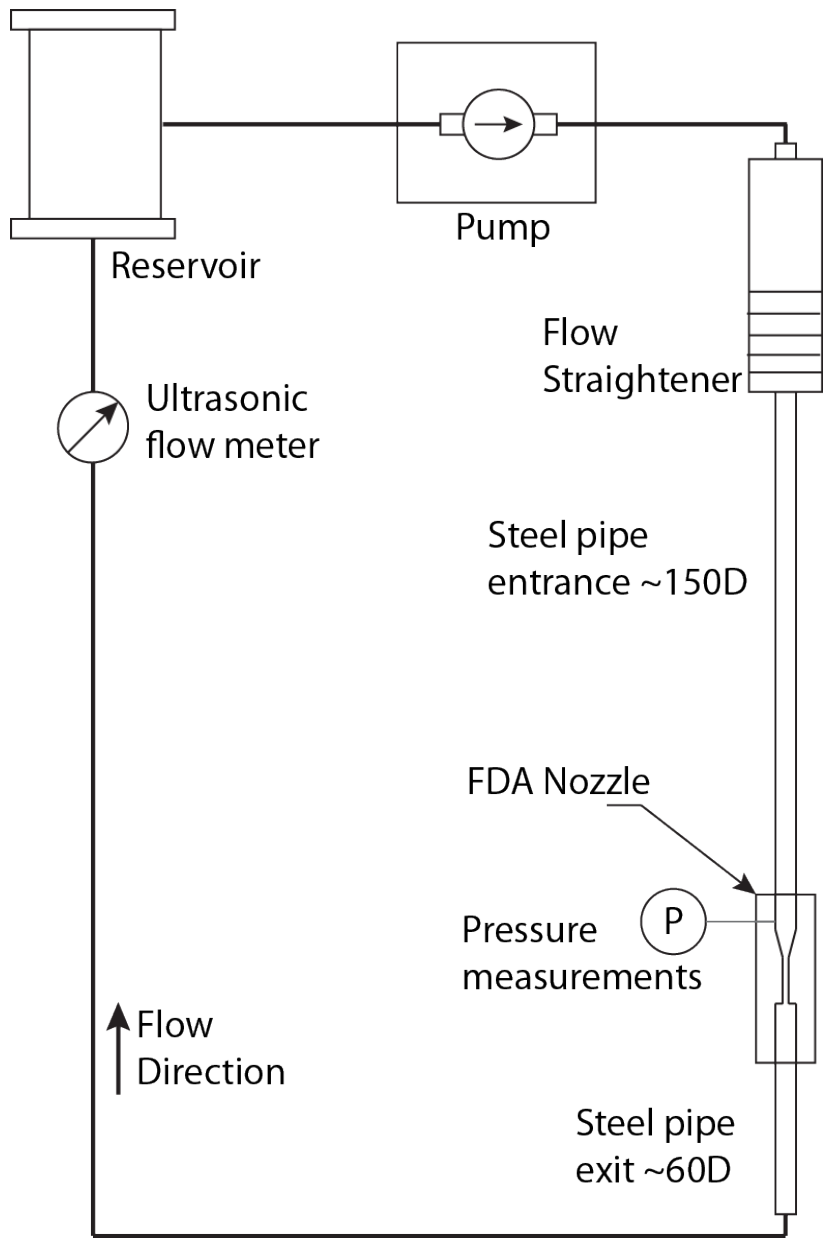

Fig. 2. Schematic of flow facility used to acquire PIV data.

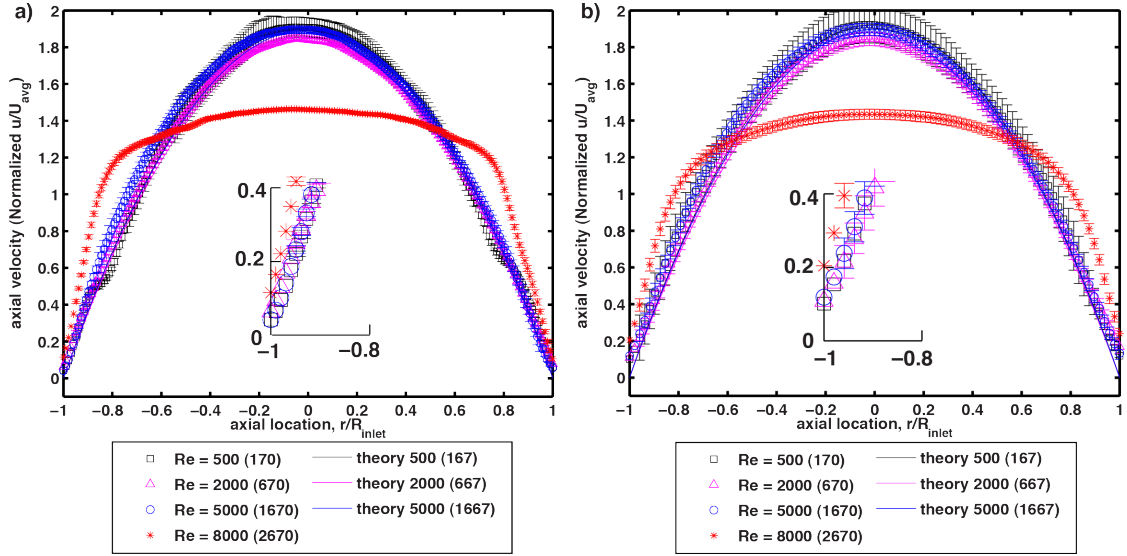

**Fig. 3.** Inlet ensemble correlation velocity profiles normalized by the inlet average velocity located at x = -21$D_{th}$. Theoretical (laminar) profiles plotted in solid lines for $Re_{th}$ =500, 2000, and 5000 ($Re_{in}$ = 170, 670, and 1670). Uncertainty bars show 95% confidence intervals. (a) ensemble correlation, (b) time-averaged velocity fields.

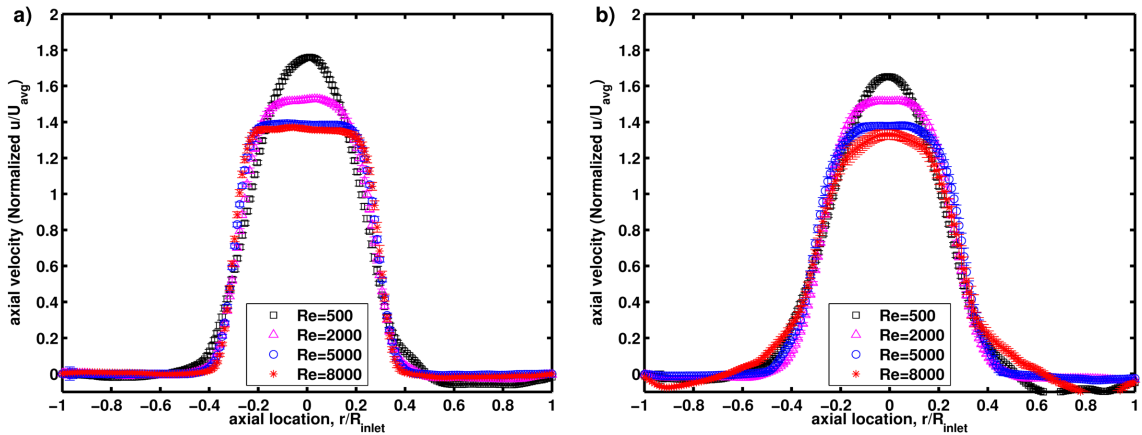

**Fig. 4.** Ensemble correlation velocity profiles normalized by the throat average velocity at x = 2$D_{th}$ (left) and x = 5$D_{th}$ (right) with respect to the sudden expansion. Uncertainty bars represent 95% confidence intervals.

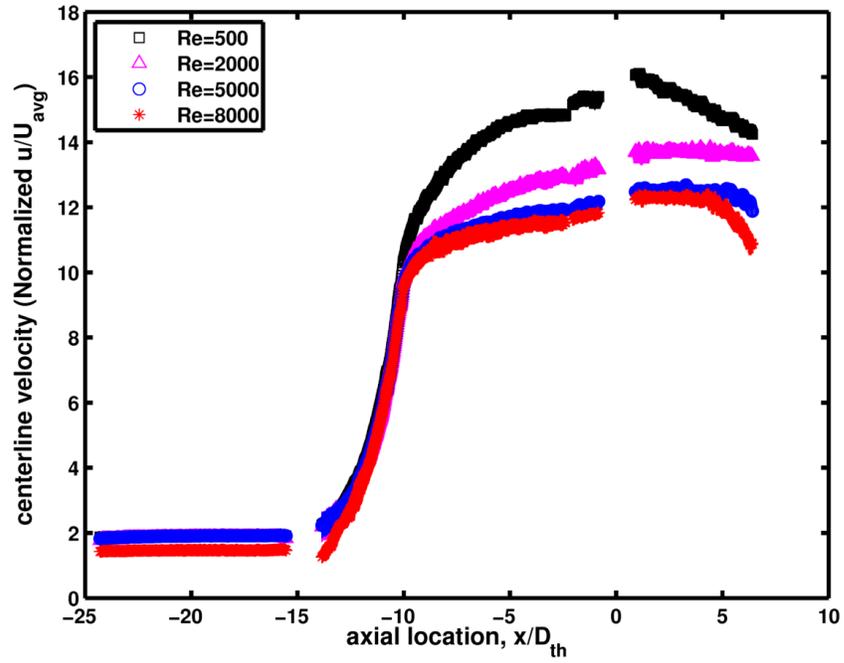

**Fig. 5.** Centerline velocities for ensemble correlation fields normalized by the average inlet velocity. Uncertainty bars indicate 95% confidence intervals.

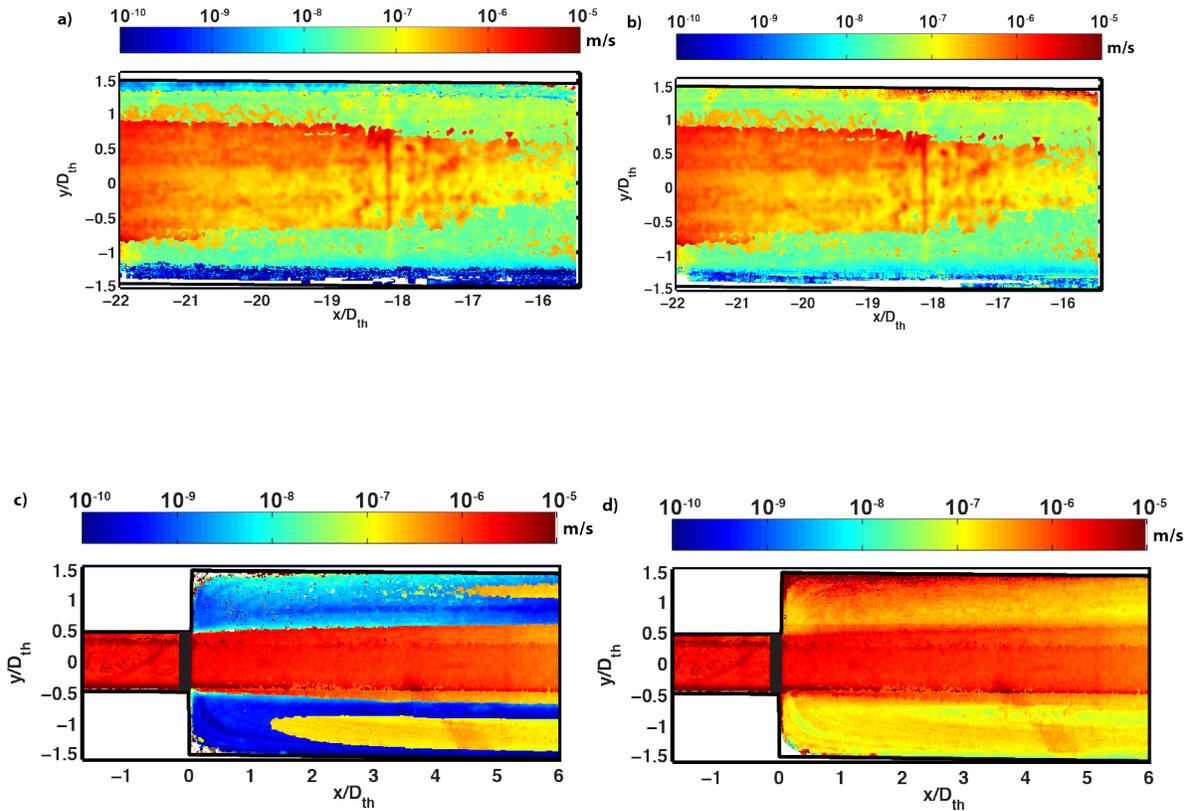

**Fig. 6.** Uncertainty field for ensemble correlation velocity, reported as absolute uncertainty in m/s for Re = 500 at (a) location 1 with dynamic range enhancement, (b) location 1 without dynamic range enhancement, (c) location 5 with dynamic range enhancement, (d) location 5 without dynamic range enhancement.

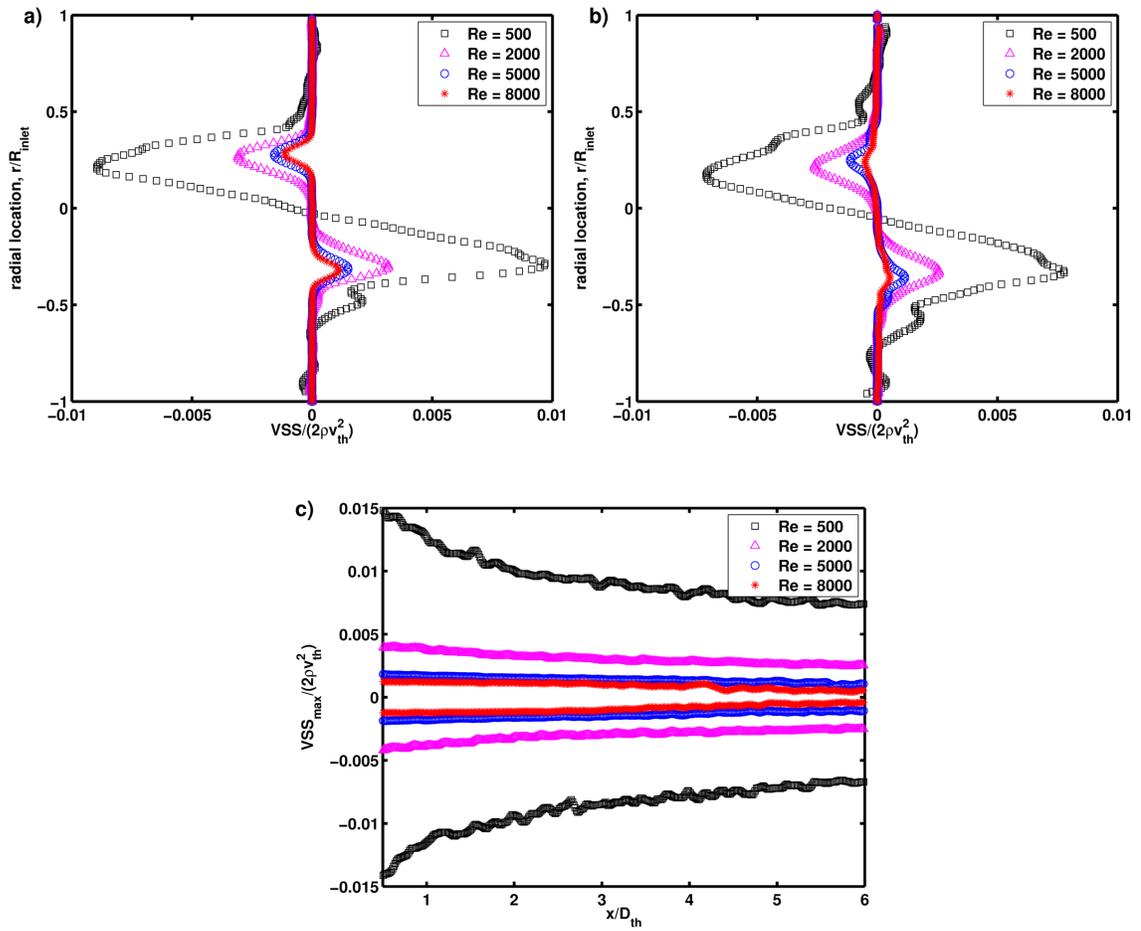

Fig. 7. Viscous Shear stress profiles normalized by $2\rho v_{th}^2$ at (a): $x = 2D_{th}$ and (b) $x = 5D_{th}$. (c): Maximum and minimum viscous shear stress profiles normalized by $2\rho v_{th}^2$ downstream of the sudden expansion. All plots derived from ensemble correlation fields.

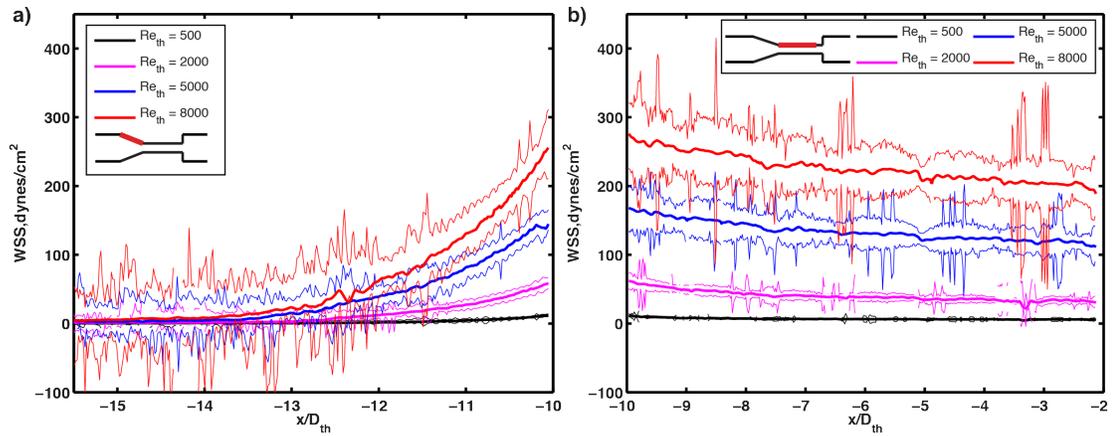

**Fig. 8 Wall shear stress computed at (a) the contraction and (b) throat walls. Thick lines represent computed WSS for each Reynolds number and thin lines represent 95% confidence intervals for the local measurement.**

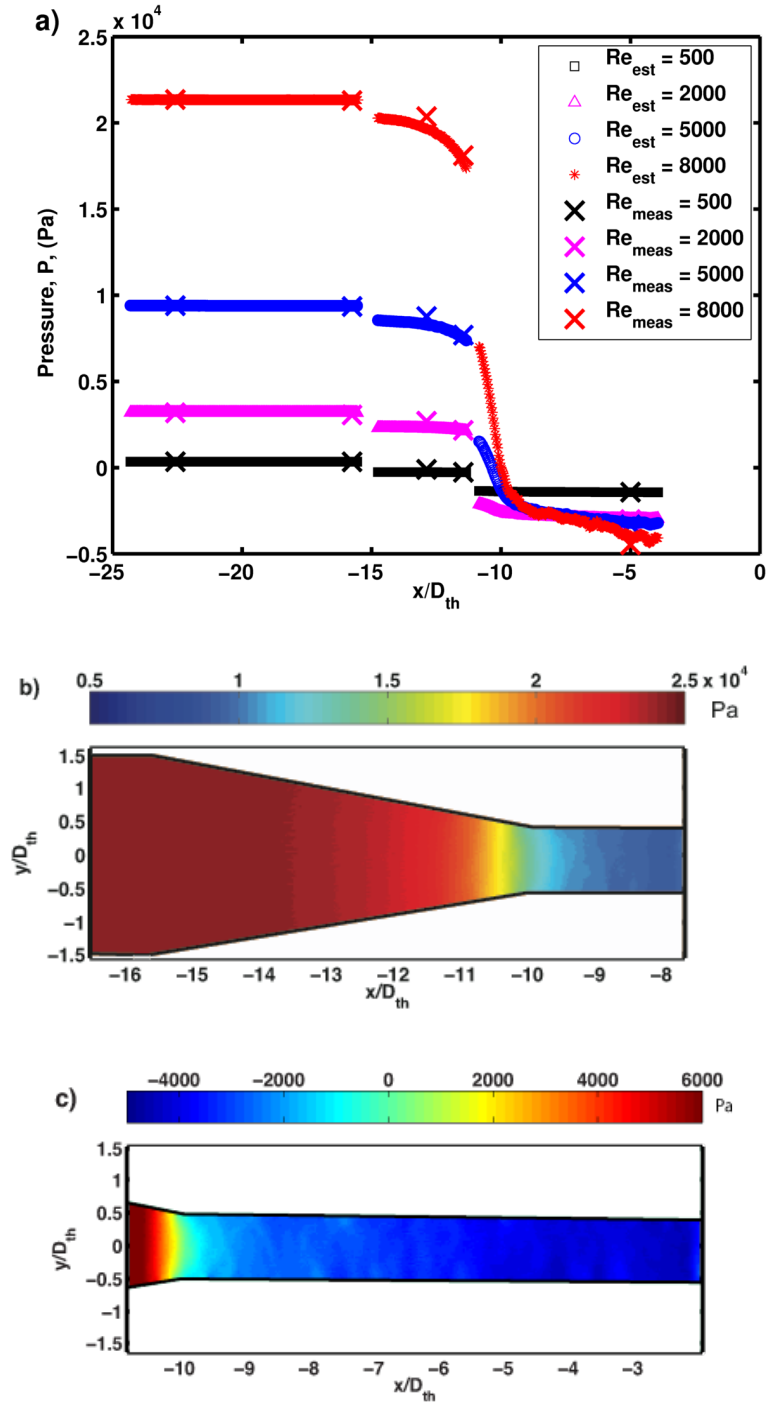

Fig. 9. (a) Pressure measurements at five tap locations shown in colored 'X' symbols. Estimated centerline pressures from relative pressure calculation plotted on same figure. Pressure fields for Re 8000 computed from instantaneous PIV fields at (b) location 2 and (c) location 3.

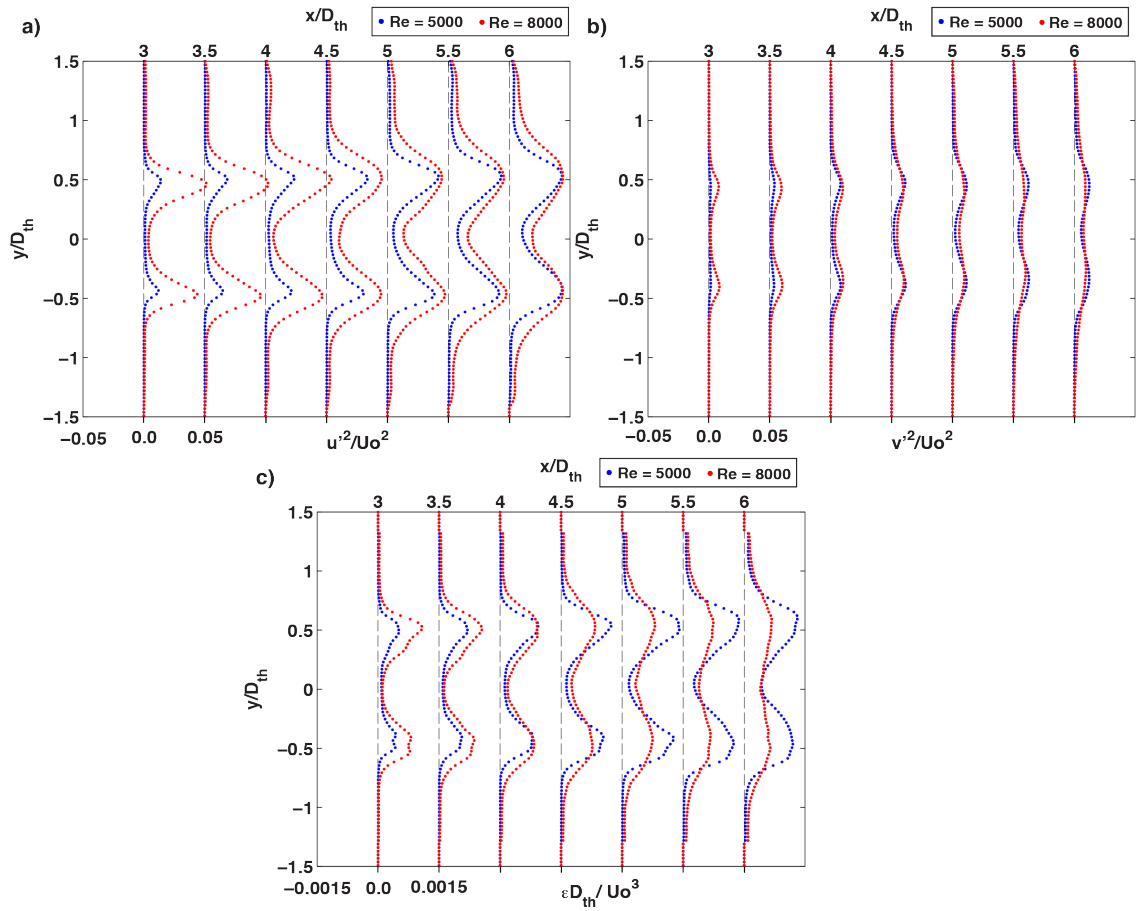

**Fig. 10.** Reynolds stresses downstream of the throat in the sudden expansion region plotted for (a) streamwise and (b) spanwise velocity components and (c) Five term energy dissipation rate for throat Reynolds numbers 5000 and 8000. Plots derived from instantaneous processing.

**Table 1. Symmetry indices for the ensemble correlation velocity profiles for each Reynolds number**

| Reynolds throat (Reynolds inlet) | Symmetry Index Ensemble correlation | Symmetry Index time averaged field |
|---|---|---|
| 500 (170) | 0.99 | 0.97 |
| 2,000 (670) | 0.96 | 0.99 |
| 5,000 (1,670) | 0.90 | 0.94 |
| 8,000 (2,670) | 0.93 | 0.99 |

**Table 2. Percentiles (5, 50, 95) for velocity uncertainty normalized with inlet average velocity. Inlet average velocity was computed from the measured flow rate.**

| Re | Location 1 | Location 2 | Location 3 | Location 4 | Location 5 |
|---|---|---|---|---|---|
| 500 | 0.004, 0.046, | 0.006, 0.084, | 0.098, 0.145, | 0.003, 0.115, | 0.006, 0.074, |
| 2,00 | 0.008, 0.011, | 0.059, 0.082, | 0.133, 0.22, | 0.001, 0.119, | 0.002, 0.08, |
| 5,00 | 0.011, 0.021, | 0.066, 0.09, | 0.097, 0.111, | 0.002, 0.093, | 0.002, 0.079, |
| 8,00 | 0.008, 0.012, | 0.082, 0.115, | 0.162, 0.235, | 0.001, 0.084, | 0.054, 0.119, |

**Table 3. Median uncertainty of WSS in dyne/cm$^2$. The uncertainties after omitting contributions from velocity are shown in parentheses.**

|  | Location 1 (inlet) | Location 2 (contraction) | Location 3 (throat) | Location 4 (throat) | Location 5 (expansion) |
|---|---|---|---|---|---|
| Re$_{th}$ 500 | 0.799 (0.009) | 1.948 (0.020) | 1.303 (0.245) | 5.471 (0.691) | 3.253 (0.007) |
| Re$_{th}$ 2000 | 0.959 (0.040) | 0.845 (0.011) | 6.209 (1.151) | 25.049 (3.682) | 8.174 (0.027) |
| Re$_{th}$ 5000 | 3.076 (0.098) | 18.064 (0.331) | 21.271 (2.272) | 42.150 (5.269) | 13.220 (0.012) |
| Re$_{th}$ 8000 | 3.188 (0.248) | 30.697 (0.465) | 36.325 (3.992) | 61.164 (8.303) | 14.433 (0.005) |